\documentclass[prl,twocolumn,superscriptaddress,showpacs,showkeys]{revtex4}
\usepackage{graphics}
\usepackage{amsmath}
\usepackage{amssymb}
\usepackage{amsfonts}
\usepackage{hyperref}

\include{MyMc} 

\renewcommand{\section}[1]{}

\begin{document}


\title{
  Controllable plasma energy bands in a 1D crystal of fractional Josephson vortices
}

\author{H.~Susanto}
\email{h.susanto@math.utwente.nl}
\affiliation{Department of Applied Mathematics, University of
Twente, P.O. Box 217, 7500 AE Enschede, The Netherlands }

\author{E.~Goldobin}
\email{gold@uni-tuebingen.de}
\author{D.~Koelle}
\author{R.~Kleiner}
\affiliation{
  Physikalisches Institut --- Experimentalphysik II,
  Universit\"at T\"ubingen,
  Auf der Morgenstelle 14,
  D-72076 T\"ubingen, Germany
}

\author{S. A.~van Gils}
\affiliation{Department of Applied Mathematics, University of
Twente, P.O. Box 217, 7500 AE Enschede, The Netherlands }




\date{\today}

\begin{abstract}

We consider a 1D chain of fractional vortices in a long Josephson junction with alternating $\pm\kappa$ phase discontinuities. Since each vortex has its own eigenfrequency, the inter-vortex coupling results in eigenmode splitting and in the formation of an oscillatory energy band for plasma waves. The band structure can be controlled at the design time by choosing the distance between vortices or \emph{during experiment} by varying the topological charge of vortices or the bias current. Thus one can construct an artificial vortex crystal with controllable energy bands for plasmons.

\end{abstract}

\maketitle

\section{Introduction}


The study of crystals is a cornerstone of solid state physics\cite{Kittel:IntroSSP,AshcroftMermin:SSP,IntroCrystallography}. The electronic structure of crystals, such as energy bands and the dispersion relation for electrons and phonons, has a great impact on all physical properties of solids. A crystal is a periodic arrangement of a group of atoms translated in space by direct lattice vectors (usually edge effects are negligible). This periodicity in space results in the formation of energy bands and different propagation modes for electrons and phonons. If one could control the periodicity, \eg by varying the mutual position and the kinds of atoms in a crystal, during experiment, one could turn a metal into an insulator and basically create a material with any desired properties. Unfortunately, the crystal structure (the type of ions, lattice and spacing) is fixed by nature and can be changed only a little, \eg by applying strong pressure, electric or magnetic fields.

Therefore, it is interesting to construct artificial periodic structures (crystals), whose properties \emph{can be varied over a wide range during experiment}, allowing a great degree of control over their resulting electronic properties. Such structures must not necessarily consist of atoms, but could consist of other more macroscopic objects which can be fabricated and arranged by means of modern technology, \eg, using lithography. 

In this letter we consider an artificial 1D crystal made of fractional Josephson vortices (fractional magnetic flux quanta), and study its plasma energy bands, \ie, energy bands of plasmons --- small oscillations of the Josephson phase.

Usually, vortices in Josephson junctions and superconductors (S) carry one quantum of magnetic flux $\Phi_0\approx2.07\times 10^{-15}\units{Wb}$.
Fractional Josephson vortices spontaneously appear in long Josephson junctions (LJJs) with an unusual current-phase relation\cite{Mints:2002:SplinteredVortices@GB,Buzdin:2003:phi-LJJ}, with spatially alternating sign of the critical current\cite{Bulaevskii:0-pi-LJJ} (\eg, SFS LJJs\cite{Ryazanov:2001:SFS-PiJJ,Kontos:2002:SIFS-PiJJ} with a step-wise variation of the ferromagnetic (F) layer thickness) or with $\pi$ discontinuities of the Josephson phase\cite{Goldobin:SF-Shape} (\eg, $d_{x^2-y^2}$-wave based LJJs\cite{Kirtley:SF:HTSGB,Kirtley:SF:T-dep,Tsuei:Review,VanHarlingen:1995:Review,Sugimoto:TriCrystal:SF,Smilde:ZigzagPRL,Hilgenkamp:zigzag:SF,Lombardi:2002:dWaveGB}). More general, a $\kappa$-discontinuity of the phase at $x=0$ ($x$ is the coordinate along the LJJ) means that the relation between the supercurrent $I_s$ and the phase $\mu$ (the first Josephson relation) in the region $x<0$ is $I_s=I_c\sin(\mu)$, while in the region $x>0$ it is $I_s=I_c\sin(\mu+\kappa)$. Fractional vortices spontaneously appear to compensate the $\kappa$-phase jump and are pinned at the discontinuity points\cite{Goldobin:2KappaGroundStates}. For a $\pi$ discontinuity the vortex carries the fractional flux $\Phi_0/2$ and the junction can be described by the same model as the junction with alternating sign of critical current\cite{Goldobin:SF-Shape,Xu:SF-shape,Kirtley:IcH-PiLJJ,Bulaevskii:0-pi-LJJ,Kogan:3CrystalVortices,Kuklov:1995:Current0piLJJ} mentioned above.

Recently, a LJJ with artificial phase discontinuities was proposed, implemented and successfully tested\cite{Goldobin:Art-0-pi}. In this junction, made using standard Nb-AlO$_x$-Nb technology, the discontinuity of the phase is created using a pair of tiny current injectors attached to the same electrode of the junction as close as possible to each other. By passing the current $I_{\rm inj}$ from one injector to the other, one can create an arbitrary $\kappa$-discontinuity of the Josephson phase with $\kappa \propto I_{\rm inj}$. Since the Josephson phase is $2\pi$-periodic, without loss of generality we consider only $0 \leq \kappa <2\pi$. One can also fabricate such a junction with as many injector pairs as required and tune the strength $\kappa$ of \emph{all} discontinuities by a single control current. Nb-AlO$_x$-Nb junctions also have very low damping (McCumber-Stewart parameter $\beta_c \gg 1$), which makes them a perfect candidate to study propagation of plasma waves and formation of energy bands. In comparison, SFS and $d$-wave based LJJs typically have  $\beta_c \lesssim 9$ (see, \eg, Fig.~2 of Ref.~\onlinecite{Smilde:ZigzagPRL}).

A fractional vortex has an eigenfrequency $\omega_0(\kappa)$ which corresponds to oscillations of the magnetic flux around the discontinuity point\cite{Goldobin:2KappaEigenModes}. This eigenfrequency lays within the plasma gap and depends on the flux carried by the vortex\cite{Goldobin:2KappaEigenModes}. If one considers two vortices at some distance from each other, so that their magnetic fields and supercurrents overlap, then the vortices behave as two coupled oscillators: their eigenfrequency splits into two frequencies, corresponding to two different modes: in-phase and out-of-phase. This splitting of eigenfrequencies for different two-vortex-configurations was already investigated numerically\cite{Goldobin:2KappaEigenModes}. In an infinite 1D array of fractional vortices situated not very far from each other, due to coupling, the eigenfrequencies split and form an energy band, very similar to the formation of bands in a crystal starting from the discrete energy levels of electrons in a single atom. In this paper we study the energy bands corresponding to oscillations of a chain of fractional vortices and we show that the band structure can be easily controlled during experiment.

\section{Model and Numerics}

Consider an infinite array of alternating $\pm\kappa$ discontinuities at a distance $a$ from each other and $\mp\kappa$ vortices pinned at them. Such an antiferromagnetically (AFM) ordered state represents the most natural configuration or, in other words, the ground state of the system. 

To model an infinite chain of vortices we consider an annular LJJ with two phase discontinuities placed at the same distance $a$ from each other, so that $2a=L$, where $L$ is the length (circumference) of the LJJ. It is also possible to realize such a geometry experimentally using an annular LJJ with injectors\cite{Goldobin:Art-0-pi,Ustinov:2002:ALJJ:InsFluxon,Malomed:2004:ALJJ:Ic(Iinj)}. The dynamics of the Josephson phase is described by the following perturbed sine-Gordon equation\cite{Bulaevskii:0-pi-LJJ,Xu:SF-shape,Goldobin:SF-Shape}
\begin{equation}
  \mu_{xx}-\mu_{tt}=\sin\left[\mu+\theta(x)\right]-\gamma,\\
  \label{eq:sg}
\end{equation}
with
\begin{equation}
  \theta(x)=\left\{
    \begin{array}{rl}
      0,      &\quad 0<x<a,\\
      -\kappa,&\quad a<x<2a=L.
    \end{array}\right.
  \label{eq:theta}
\end{equation}
The Josephson phase $\mu(x,t)$ is a continuous\cite{EndNote:Phases} function of the spatial variable $x$ and time $t$,  normalized to the Josephson penetration depth $\lambda_J$ and to the inverse plasma frequency $\omega_p^{-1}$, respectively. The subscripts $x$ and $t$ denote derivatives with respect to space and time. We normalize the bias current density $j$ to the critical current density $j_c$, \ie, $\gamma\equiv j/j_c$. The function $\theta(x)$ defines the positions and the strength of the discontinuities. Since we consider an annular geometry we supplement Eq.~(\ref{eq:sg}) with the periodic boundary conditions: $\mu(0,t)=\mu(L,t)$ and $\mu_x(0,t)=\mu_x(L,t)$.

To calculate numerically the oscillatory energy bands of a chain of fractional vortices, we, first of all, calculate the steady solution corresponding to an AFM vortex chain for given values of $\kappa$, $a$, and $\gamma$ by solving Eq.~(\ref{eq:sg}) numerically\cite{This:numerics}. We use several spatial discretization steps to compare the obtained results, \ie, $\Delta x=0.1,\,0.05,\,0.02$.

If $\mu_0(x)$ is the obtained static solution of Eq.~(\ref{eq:sg}), then the small oscillations around this solution are given by
\begin{equation}
  \mu(x,t) = \mu_0(x) + \epsilon(x) e^{i\omega t}
  . \label{Eq:SmallOscillations}
\end{equation}
Substituting expression (\ref{Eq:SmallOscillations}) in Eq.~(\ref{eq:sg}) we get 
\begin{equation}
  \epsilon_{xx} + \omega^2 \epsilon = \cos[\mu_0(x)+\theta(x)]\epsilon
  . \label{Eq:EigenvalueProblem}
\end{equation}
Written as a system of first order equations, the eigenvalue problem (\ref{Eq:EigenvalueProblem}) takes the form
\begin{equation}
  \vec{\epsilon}_x = A \vec{\epsilon}
  ,\quad\text{where}\quad \vec{\epsilon}=
  \left( \begin{array}{l}\epsilon \\ \epsilon_x\end{array}\right)
  \label{eq:evp}
\end{equation}
with the matrix 
\begin{equation}
  A(x)=
  \begin{pmatrix}
    0 & 1\\ 
    \cos\left[\mu_0(x)+\theta(x)\right]-\omega^2 & 0
  \end{pmatrix}.
  \label{eq:a}
\end{equation}

In our case, since $A(x)\neq const$, we cannot implicitly integrate Eq.~(\ref{eq:evp}). Instead, we approximate $A(x)$ by a constant matrix $A_n$ given by Eq.~(\ref{eq:a}) with $x=x_n=n\Delta x$ on each interval $\Delta x$ from $x_n$ to $x_{n+1}$. Then we solve Eq.~(\ref{eq:evp}) on each small interval $\Delta x$ to find $\vec{\epsilon}(x_{n+1})=\vec{\epsilon}(x_n)\exp(\Delta x A_n)$.

As a consequence of the Floquet (Bloch) theorem the solution of Eq.~(\ref{eq:evp}) should satisfy the following equation 
\[
  \vec{\epsilon}(x+2a) = C  \vec{\epsilon}(x),
\]
with constant matrix $C$, where $2a$ is the periodicity of the lattice. 
In our case, the so-called principal matrix $C$ is approximated by a simple transfer matrix discretization described above, \ie,  
\begin{equation}
  C
  =\prod_{n=1}^N{\exp(\Delta x\, A_n)}
  ,\label{Eq:C}
\end{equation}
where $N=2a/\Delta x$. More delicate schemes to calculate a band-gap can be found in, \eg, Refs.~\cite{Mendez:TM,Jazar:Scheme}. The transfer matrix $C$ has two eigenvalues. The product of these eigenvalues 
\[
  \lambda_1\lambda_2=\det C
  =\prod_{n=1}^N\det\,e^{\Delta x\, A_n}
  =\prod_{n=1}^Ne^{\Tr(\Delta x\, A_n)}=1,
\]
where we used an identity $\det[\exp(M)]=\exp[\Tr(M)]$ and the fact that $\Tr(A)=0$, see Eq.~(\ref{eq:a}). The periodic solution exists, \ie, $\omega$ is within the energy band, only if $|\lambda_1|=|\lambda_2|=1$, \ie, if the determinant of the characteristic equation is negative and $\lambda_{1,2}$ is a pair of complex-conjugate roots laying at the unit circle on the complex plane.

\section{Results}

\begin{figure*}[tb]
  \begin{center}
    \includegraphics{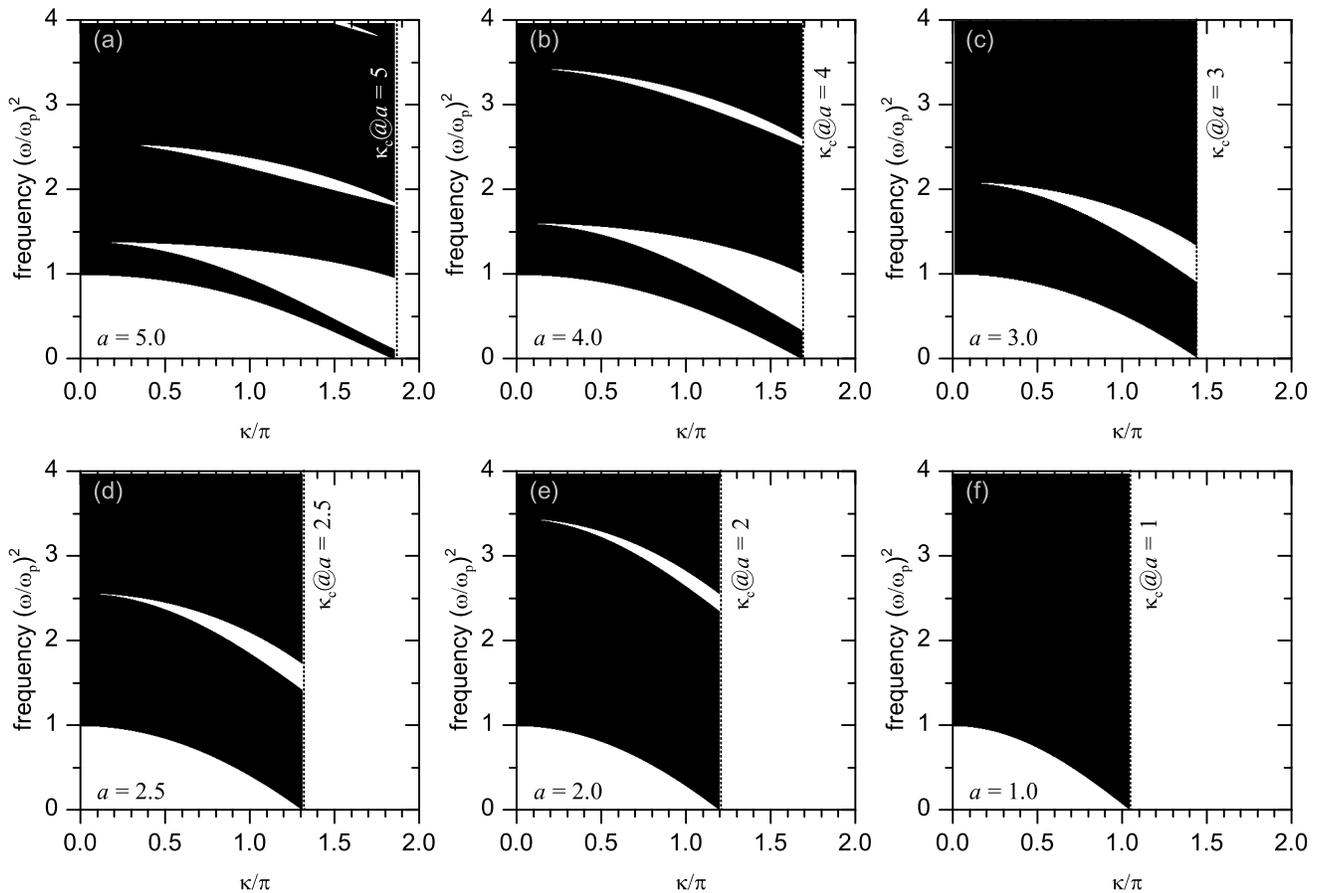}
  \end{center}
  \caption{%
    Numerically calculated band structure as a function of $\kappa$ for  $a=5,\,4,\,3,\,2.5,\,2,\,1$ shown in (a)--(f), respectively. At $\kappa=\kappa_c(a)$ a crystal becomes unstable and structural phase transition takes place.
  }
  \label{Fig:AFM:bands}
\end{figure*}

First, we consider the case when the bias current is absent, \ie, $\gamma=0$. Band structures as a function of $\kappa$, numerically calculated for different $a$ (given in units normalized to $\lambda_J$) from $a=5$ (weak coupling) to $a=1$ (strong coupling) are shown in Fig.~\ref{Fig:AFM:bands}. In all plots one can see that in the absence of discontinuities ($\kappa=0$) the junction has a plasma gap for $0<\omega<1$  ($\omega$ is normalized to the plasma frequency $\omega_p$) and a single infinite plasma band for $\omega>1$. As $\kappa$ increases, fractional vortices appear. Each vortex, if isolated, has an eigenfrequency $\omega_0(\kappa)<1$\cite{Goldobin:2KappaEigenModes}. In our case vortices are coupled and the eigenfrequency splits into a band, which is the lowest energy band, see Fig.~\ref{Fig:AFM:bands}(a). Small gaps also appear in the former continuous plasma band $\omega>1$. As the coupling increases (distance $a$ decreases), the bands broaden, while the gaps shrink and shift to higher frequencies, as can be seen in consecutive Figs.~\ref{Fig:AFM:bands}(a)--(f). 

In Fig.~\ref{Fig:AFM:bands} the bands are traced from $\kappa=0$ to $\kappa=\kappa_c(a)$. At $\kappa=\kappa_c(a)$ the AFM state becomes unstable and turns itself into a complimentary state, \ie, each of the $\pm\kappa$-vortices becomes a $\pm(\kappa-2\pi)$-vortex\cite{Goldobin:2KappaGroundStates}. As a sign of this instability the lower band touches zero, \ie, $\omega\to0$ at $\kappa\to\kappa_c$. For the complimentary state, the band structure is the mirror reflection of the one shown in Fig.~\ref{Fig:AFM:bands} with respect to the line $\kappa=\pi$. Note, that in the interval of $\kappa$ from $2\pi-\kappa_c(a)$ to $\kappa_c(a)$ there are two stable solutions: an AFM chain of direct vortices or an AFM chain of complementary vortices. Similar behavior was reported for a system of only two vortices\cite{Goldobin:2KappaGroundStates}. The value of $\kappa_c(a)$ decreases as the coupling increases reaching the value $\kappa_c\to\pi$ at $a \to 0$, in agreement with the known result\cite{Goldobin:SF-ReArrange,Zenchuk:2003:AnalXover} that the infinite AFM ordered \emph{semifluxon} chain is stable for any $a\to0$.

Contrary to a crystal of integer vortices\cite{Lebwohl:1967:JVortexLinesProp,Fetter:1968:FluctLJJ,Takayama:1993:KinkCrystal:BosonExcit}, our 1D crystal has \emph{no acoustic branch} (except the case $\kappa=2\pi n$) in the dispersion relation because $\kappa$-vortices are pinned and cannot move all together, \ie, with $\omega=0$ and $k=0$. 


%
\begin{figure*}[tb]
  \begin{center} 
   \includegraphics{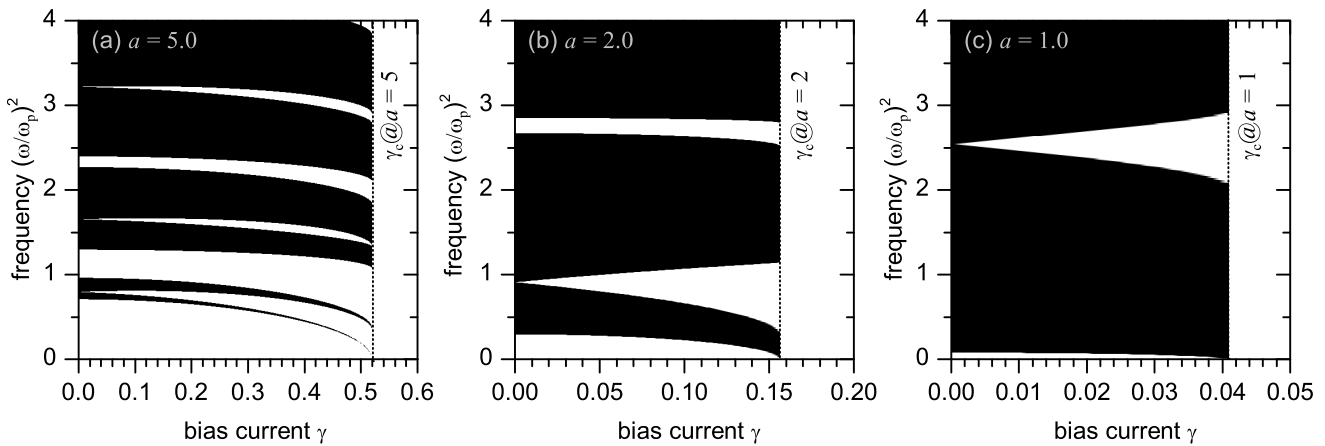}
  \end{center}
  \caption{%
    Numerically calculated band structure as a function of the applied bias current $\gamma$ for $\kappa=\pi$ and (a) weak coupling $a=5$, (b) moderate coupling $a=2$ and (c) strong coupling $a=1$. At $\gamma=\gamma_c(a,\kappa)$ a crystal becomes unstable and structural phase transition takes place.
  }
  \label{Fig:AFM:bands_gnonzero}
\end{figure*}

Next, we consider the control of the band structure by the bias current at various fixed values of $a$ and $\kappa=\pi$. This case is interesting for 0-$\pi$-LJJs where the discontinuities cannot be controlled during experiment. In Fig.~\ref{Fig:AFM:bands_gnonzero} we present the band structure for the case of weak ($a=5$), moderate ($a=2$) and strong ($a=1$) coupling. As one can see, by applying a bias current, an additional gap opens within each band. For the case of moderate and strong coupling these gaps can become quite large near the critical value of $\gamma=\gamma_c(a)$ at which the static solution becomes unstable. Again, as a sign of instability the lowest band has $\omega\to0$ at $\gamma\to\gamma_c(a)$. For $\gamma>\gamma_c(a)$ the system switches to the finite voltage state. Note, that $\gamma_c(a)\to0$ at $a\to0$ in qualitative agreement with the results for finite semifluxon chains.\cite{Goldobin:SF-ReArrange}

\section{Conclusions}

To conclude, we have calculated the energy bands corresponding to small oscillations (plasma waves) of the 1D AFM ordered fractional vortex crystal  as a function of the discontinuity $\kappa$. Such a 1D vortex crystal has no acoustic, but only an optical branch in the dispersion relation, which is a direct consequence of vortex pinning. The band structure can be changed by changing $\kappa$. In case of discontinuities created artificially by injectors, one can make a wiring such that a single control current changes the value $\kappa$ for \emph{all} discontinuities in the same time, thus providing the possibility to change the band structure ``on-the-fly''. For natural 0-$\pi$-LJJs\cite{Smilde:ZigzagPRL,Hilgenkamp:zigzag:SF,Lombardi:2002:dWaveGB}, where the discontinuity $\kappa=\pi$ is fixed, the band structure can be smoothly controlled \emph{during experiment} by the bias current. 

The knowledge of eigenmodes and the band structure is a key element in designing classical or quantum devices based on fractional vortices. In the classical domain this may help to avoid resonance phenomena or to exploit them (\eg in filters and detectors). In the quantum domain, thanks to the absence of acoustic branch, the lowest energy gap can be a crucial obstacle for thermal excitation of plasmons. 

Here, we have calculated the plasmon spectrum for a mirror symmetric crystal. However, the most unusual properties can be expected from systems with broken reflection symmetry (ratchets)\cite{APA:SpecRatchets,Reimann:2002:BrownianMotors}, such as crystals of ferroelectrics or of some superconductors.\cite{Bulaevskii:SC:AsymmCrystals} From this point of view, the transport in such systems is not well studied. Using an array of discontinuities of different strengths and distances, one might realize \emph{controllable} fractional vortex crystals without reflection symmetry and study the nonequilibrium transport. In this case the eigenvalue problem (\ref{Eq:EigenvalueProblem}) corresponds to the motion of a plasmon in a ratchet potential. 

We thank B. Malomed, L. Bulaevskii and A. Doelman for discussions. This work was supported by the Deutsche Forschungsgemeinschaft (GO-1106/1), by the ESF programs ``Vortex'' and ``Pi-shift'' and by the Royal Netherlands Academy of Arts and Sciences (KNAW).

\bibliographystyle{apsprl}
\bibliography{this,LJJ,SF,pi,SFS,software,ratch}

\begin{thebibliography}{39}
\bibitem[{End()}]{EndNote:Phases}
\bibinfo{note}{The discontinuous phase $\phi(x,t)=\mu(x,t)+\theta(x)$, which is
  the real phase in Nb-AlO$_x$-Nb junctions, is not used in this manuscript.
  See Ref.~\onlinecite{Goldobin:SF-Shape} for details.}
\end{thebibliography}

\end{document}